\documentstyle[12pt,epsfig]{article}
\title{$\Delta $-production in
$\bar{{\rm p}}$d-annihilation at rest
\thanks{Supported in part by BMBF
\newline {\it Talk given at Nucleon-Antinucleon 
International Conference NAN'95, 
Moscow, 1995, to be published in Sov. J. Nucl. Phys.} } }
\author{Amand Faessler$^1$, A. Sibirtsev$^2$ and 
K. Tsushima$^3$ \thanks{Supported by Australian Research Council
\newline Adelaide University, ADP-96-8/T213} \\ \\
{\small  $^1$Institut f\"ur Theoretische Physik,
Universit\"at T\"ubingen }\\
{\small Auf der Morgenstelle 14, D-72076 T\"ubingen, Germany } \\ \\
{\small $^2$Laboratory of Nuclear Problems, Joint Institute } \\
{\small of  Nuclear Research, Dubna, 141980, Russia } \\ \\
{\small $^3$Department of Physics and Mathematical Physics } \\
{\small The University of Adelaide, Australia 5005}}
\begin{document}
\date{ }
\maketitle
\begin{abstract}
We study the $\Delta$-excitation in $\overline{p}d$ annihilation at
rest. The invariant spectra
of the $\pi^+p$ and $\pi^-p$ systems selecting the protons
with momenta above 400 MeV/c are analyzed. The calculations reproduces 
reasonably the experimental data.

\end{abstract}

One of the traditional mechanisms of fast baryon production
following the antiproton annihilation is the secondary interactions of the 
pions as well as the mesonic resonances in the nuclear environment.
It is clear that an
essential indication for the contribution of the meson rescattering
is the excitation of the $\Delta$-resonances.

The search for isobar was performed in $\overline{p}d$ annihilation 
by Kalogeropoulos et al.~\cite{Kalogeropoulos}, but they found no
$\Delta$ influence in   $\pi N$-system. As was later shown 
by Voronov and Kolybasov~\cite{Voronov} analyzing the reaction
$\overline{p}d \rightarrow 2\pi^+3\pi^-p$ that the structure of
the isobar may be significantly smeared out due to  the 
contribution from  $\pi N$-pairs in which the pion does not undergo
rescattering. It was suggested that the $\Delta$-resonance 
can be observed only when the recoil nucleons are selected with
momenta above 200 MeV/c.

Recently the  OBELIX collaboration from CERN measured the
annihilation of stopped antiprotons in a
deuteron target~\cite{Ableev}.
The invariant  mass of protons with momenta larger than 400 MeV/c 
and $\pi^+$-mesons indicates a clean peak from $\Delta^{++}$-resonances,
whereas the $\Delta^0$-resonance in the $\pi^-p$ system was not found.

Here we study the $\Delta$-excitation in
$\overline{p}d$ annihilation at rest. 
  The annihilation amplitude from the 
statistical model and the $\pi N$ amplitude
from the resonance model were adopted in our calculations.

Similar to~\cite{Kolybasov1} we
calculate the amplitude of the triangle diagram shown in fig.1 as
\begin{equation}
T = \frac {E_{\pi}+E_p} {2 \pi (m-E_{\pi})}
\int \frac {T_1(\overline{p}N \rightarrow n\pi) T_2(\pi N)}
{{\bf k}_1^2 - {\bf k}_2^2+i\epsilon} \phi ({\bf k}_1+{\bf Q}/2)
d{\bf k}_1
\end{equation}
where $T_1$ is the amplitude of $\overline{p}N \rightarrow n \pi$
annihilation, $E_{\pi}$ and $E_p$ are the pion and 
the proton energies in the final state, ${\bf k}_1$ and ${\bf k}_2$
are the pion momenta before and after the interaction. Here $m$ is the nucleon
mass and ${\bf Q}$ is the deuteron momentum.
The deuteron wave function $\phi ({\bf Q})$
 for the Bonn potential~\cite{Bonn1} was adopted.

We use the annihilation probability from~\cite{Hernandez} as
\begin{equation}
\label{am1}
| T_1(\overline{p}N \rightarrow n\pi) | ^2 = 
G_f \frac {\lambda^{1/2}(s,
m^2,m^2)}{2m^2} \frac {P_n} {I_n(s)}
\end{equation}
where  $s$ is the squared invariant mass of
$\overline{p}N$ system and $\lambda$ is the K\"alen function.
Factor $G_f$ stands for the charge configuration of the final system of
$n$-pions and was calculated with the statistical approach as~\cite{Pais}
\begin{equation}
\label{stat}
G_f= \left[n_+! n_-! n_0! \right]^{-1}
 \left[ \sum_{\beta} \left(n_+! n_-! n_0! \right)_{\beta}^{-1}
\right]^{-1}
\end{equation}
where $n_+$, $ n_-$ and $ n_0$ are the numbers of positive, negative and
neutral $\pi$-mesons for a final charge system of $n=n_++n_-+n_0$
pions and the summation is performed over all reaction channels allowed for a
given $n\pi$ system.

In eq.(\ref{am1}) the factor $I_n(s)$ accounts for the phase space volume of
$n$-pions with invariant mass $\sqrt{s}$ and $P_n$ stands for the probability
for the  creation of $n$-pions in $\overline{p}N$ annihilation, which was
taken as~\cite{Stenbacka}
\begin{eqnarray}
P_n= \left(2\pi \sigma \right)^{1/2} exp \left[ -\frac {(n-\nu)^2}
{2\sigma} \right] \nonumber \\
\nu =2.65+1.78 lns, \ \ \ \sigma=0.174 \nu s^{0.2} 
\end{eqnarray}

Accordingly to Hernandez et al.~\cite{Hernandez} the annihilation amplitude
has a smooth momentum dependence and the momentum distribution of the
pions are described by the $n$-body phase space.

The $T_2(\pi N )$ is the amplitude which accounts for the
$\pi N \rightarrow \pi N$ scattering in the $\Delta$-isobar region.
This amplitude
 is calculated with the resonance model~\cite{Tsushima1,Tsushima2}.
The relevant interaction Lagrangian is given by,
\begin{equation}
{\cal L}_{\pi N \Delta} = 
\frac{g_{\pi N \Delta}}{m_\pi}
\left( \bar{\Delta}^\mu \overrightarrow{\cal I} N \cdot 
\partial_\mu \vec\phi + \bar{N} {\overrightarrow{\cal I}}^\dagger 
\Delta^\mu \cdot \partial_\mu \vec\phi \, \right),
\label{pind}
\end{equation}
where $\vec{\cal I}$ is the isospin transition operator
$\overrightarrow{\cal I}_{Mm} = \sum_{\ell=\pm1,0} 
(1 \ell \frac{1}{2} m | \frac{3}{2} M) 
\hat{e^*}_{\ell}$,  
and  $\vec \phi $ stands for the pion field. The coupling constant 
$g_{\pi N \Delta}$ appearing in the Lagrangian is determined by 
the experimental width $\Gamma_{\Delta}$ of the $\Delta$ resonance as,
\begin{equation}
\Gamma_{\Delta} =
\frac{g^2_{\pi N \Delta} F^2({\bf q})}{12\pi}
\frac{(\sqrt{m_N^2+{\bf q}^{\hspace{1mm}2}}+ m_N)}
{m_{\Delta} m_{\pi}^2} |{\bf q}|^3
\label{gpind},
\end{equation}
with $|{\bf q}|=\lambda^{\frac{1}{2}}
(m_{\Delta}^2,m_N^2,m_{\pi}^2)/(2m_{\Delta})$ and $F({\bf q})$ being
the form factor which simulates the finite size effects of the hadrons.
This form factor is multiplied to each $\pi N \Delta$ vertex in the 
calculations, and explicit expression is, 
\begin{equation}
F({\bf q}) = \frac{\Lambda^2}{\Lambda^2 + {\bf q}\,^2},
\end{equation}
with $\Lambda$ being the cut-off parameter.
For the propagator $G_{\Delta}^{\mu \nu}$ of the $\Delta$ resonance, we use
\begin{equation}
i G_{\Delta}^{\mu \nu}(p) =i \frac{-P^{\mu \nu}(p)}{p^2 - m_{\Delta}^2 +
im_{\Delta}\Gamma_{\Delta}}\,,
\end{equation}
with
\begin{equation}
P^{\mu \nu}(p) = - (\gamma \cdot p + m_{\Delta}) 
\left[ g^{\mu \nu} - \frac{1}{3} \gamma^\mu \gamma^\nu
- \frac{1}{3 m_{\Delta}}( \gamma^\mu p^\nu - \gamma^\nu p^\mu)
- \frac{2}{3 m_{\Delta}^2} p^\mu p^\nu \right].   \label{pmunu}
\end{equation}
The cut-off parameter $\Lambda$ appearing in the form factor $F({\bf q})$
was fitted to the experimental
total cross section for the $\pi^+p$ interaction~\cite{Landolt}.
We get $\Lambda$=0.33 GeV, which corresponds
to the value $g_{\pi N \Delta}^2/4\pi = 0.83$, by eq. (\ref{gpind}) 
(without $F^2(\vec q)$ in eq. (\ref{gpind}) gives this value 0.38, 
that is comparable to the value of the Bonn potential~\cite{bonn} 0.35.).

Calculated  spectra of the $\pi^+p$ invariant mass
from  the reaction 
$\overline{p}d \rightarrow p \pi^+ 2\pi^- m\pi^0$ are shown in
Fig.2. The experimental data were taken from~\cite{Ableev}.
Similar as in ref.~\cite{Ableev} we select the protons with momenta
above 400 MeV/c. The dashed line shows the $\Delta^{++}$-resonance,
whereas the dotted line shows the combinatorial background that comes
from the pions which do not undergo interactions. The solid line is the
sum and describes  reasonably the experimental data.
The isobar structure might be clearly reconstructed from the 
invariant mass distribution.

A quite different situation exists for the spectrum of the 
$\pi^-p$ invariant mass 
from the reaction $\overline{p}d \rightarrow p \pi^+ 2\pi^- m\pi^0$,
which is shown in Fig.3.
The signal from the $\Delta^0$ resonance is to weak to be extracted from
strong combinatorial background. In Fig.4 we show  separate contributions 
from $\overline{p}d \rightarrow p \pi^+ 2\pi^- \pi^0$ (a) and
$\overline{p}d \rightarrow p \pi^+ 2\pi^- 2\pi^0$ (b) reactions to
the $\pi^-p$ invariant mass. Most promising is to study the  $\Delta^0$
excitation in the reaction $\overline{p}d \rightarrow p \pi^+ 2\pi^- \pi^0$,
which shows narrow peak at the isobar region and large ratio of the
$\Delta^0$ amplitude to the combinatorial background at
$M (\pi^-p) \simeq 1.25$ GeV.

\newpage

\begin{figure}[h]
\epsfig{file=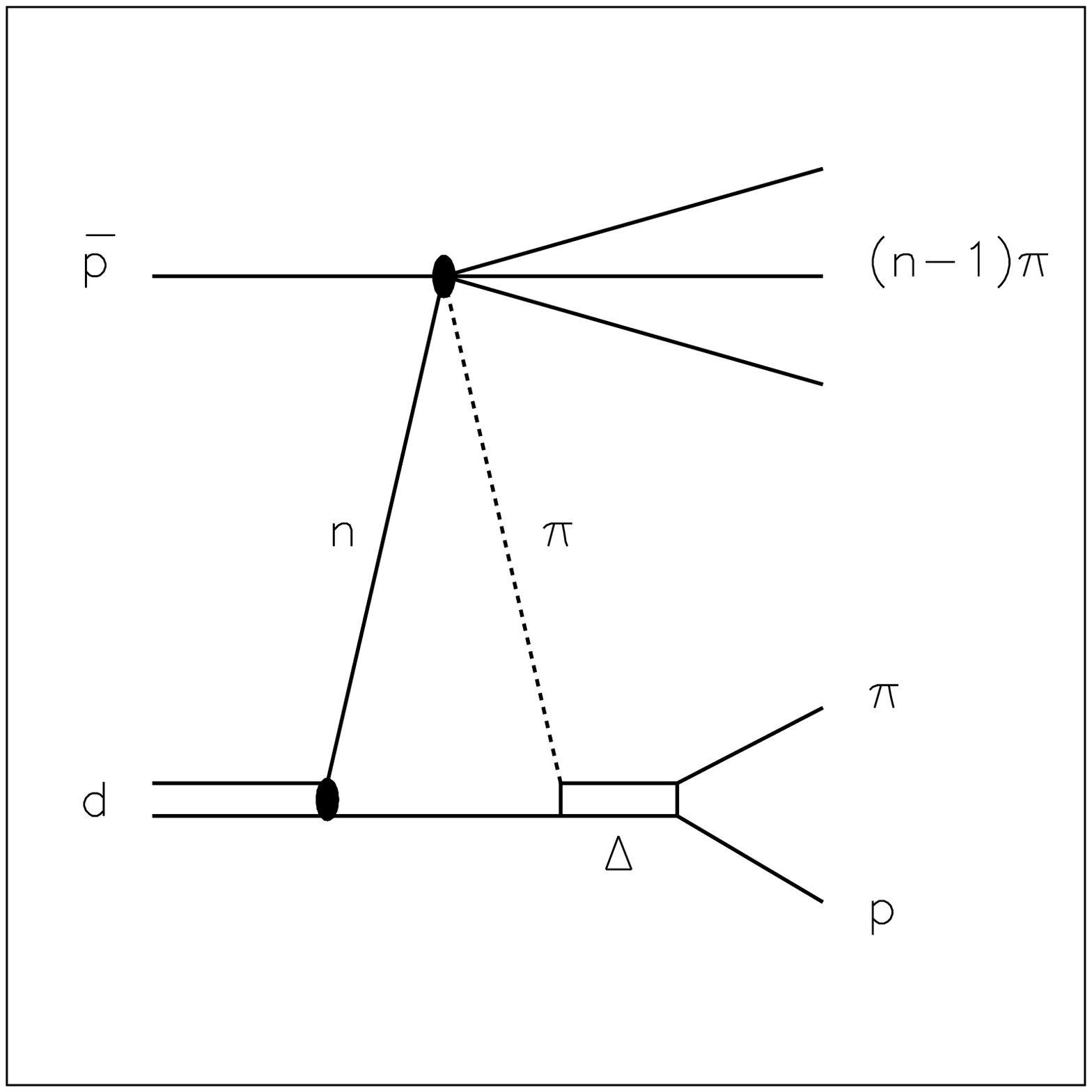,width=10cm}
\caption{Pion rescattering diagram.}
\end{figure}

\begin{figure}[h]
\epsfig{file=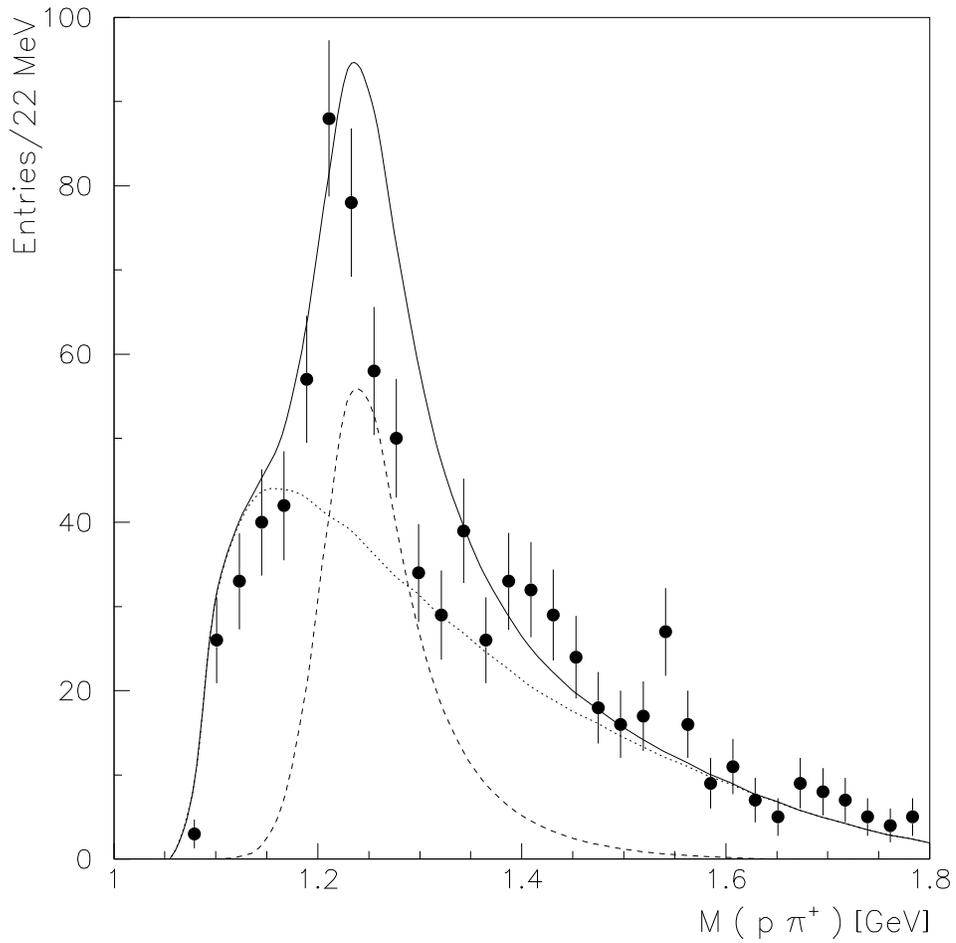,width=14cm}
\caption{The spectrum of the $\pi^+p$ invariant mass
from the reaction 
$\overline{p}d \rightarrow p \pi^+ 2\pi^- m\pi^0$ for 
protons with momenta above 400 MeV/c. Experimental data are 
from~\protect\cite{Ableev} and lines show our results. Dashed line is 
the contribution from
$\Delta^{++}$-excitation, dotted from the 
combinatorial background, while
the solid line shows the sum.}
\end{figure}

\begin{figure}[h]
\epsfig{file=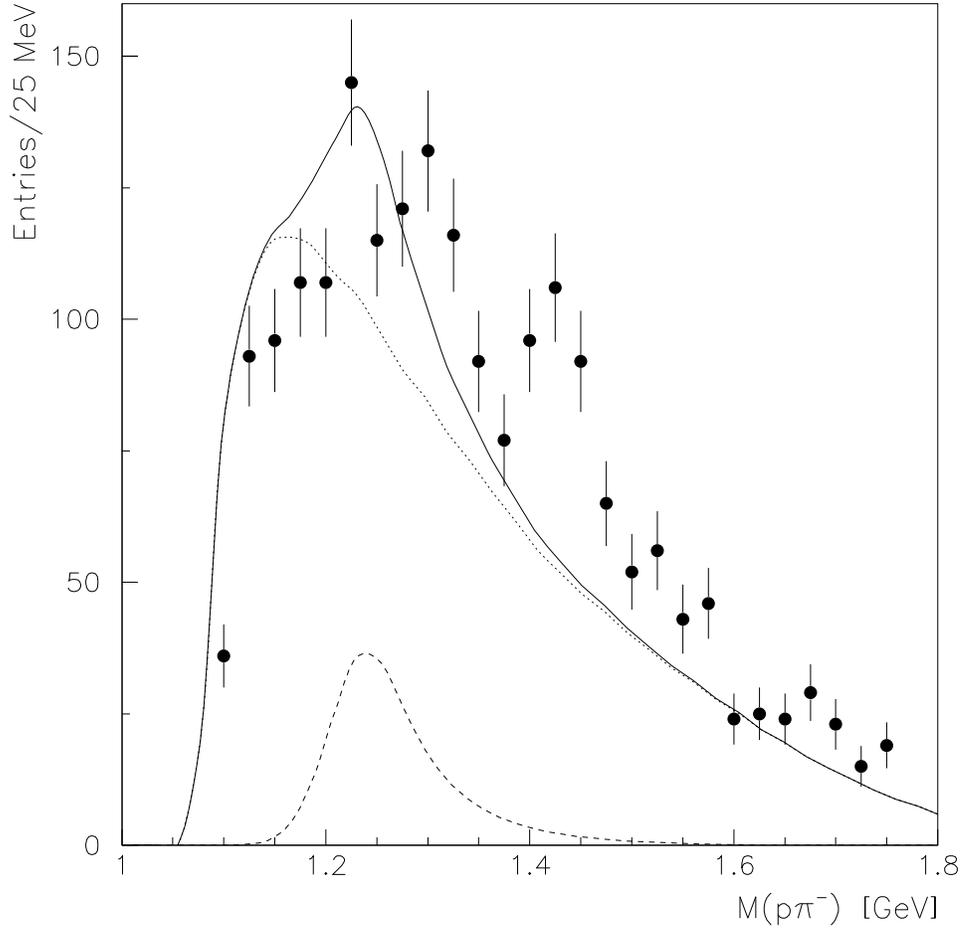,width=14cm}
\caption{The spectrum of the $\pi^-p$ invariant mass
from the reaction 
$\overline{p}d \rightarrow p \pi^+ 2\pi^- m\pi^0$ for 
protons with momenta above 400 MeV/c. Experimental data are 
from~\protect\cite{Ableev} and lines show our calculations.
Dashed line is 
the contribution from
$\Delta^{0}$-excitation and dotted from the background. Solid line
shows the sum.}
\end{figure}

\begin{figure}[h]
\epsfig{file=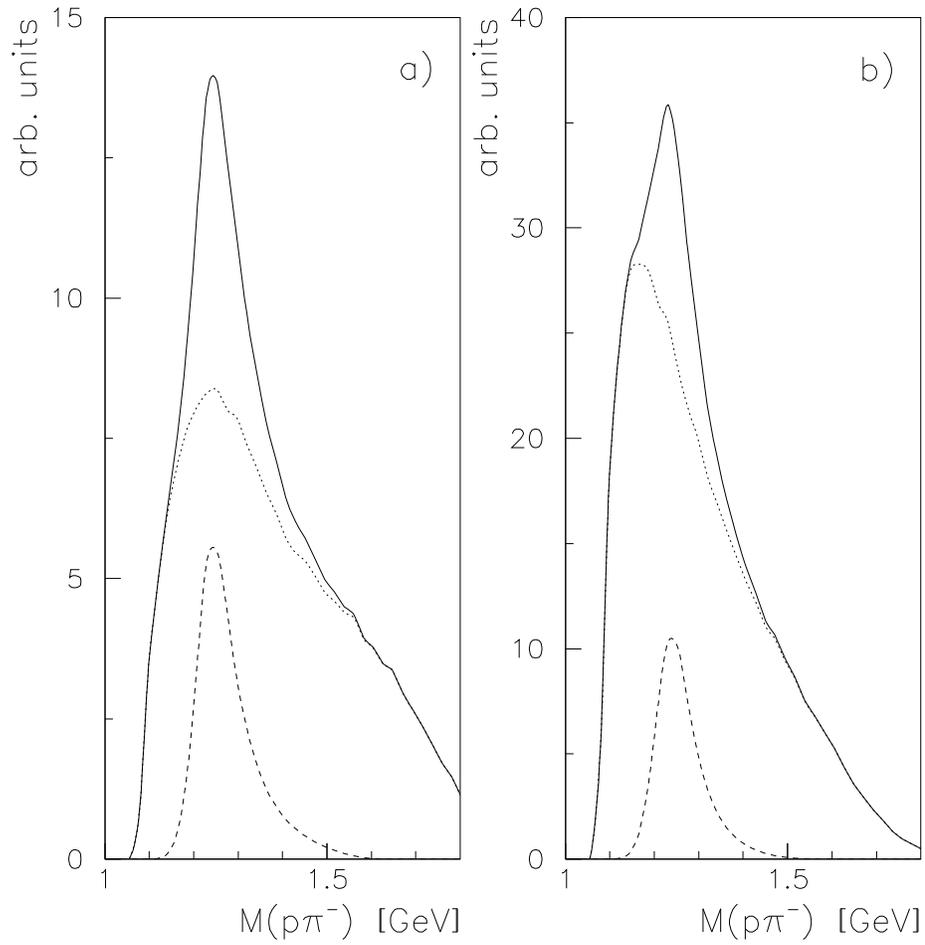,width=14cm}
\caption{The  spectra of the $\pi^-p$ invariant mass 
from the reactions 
$\overline{p}d \rightarrow p \pi^+ 2\pi^- \pi^0$ (a)
and $\overline{p}d \rightarrow p \pi^+ 2\pi^- 2\pi^0$ (b)
and for protons with momenta above 400 MeV/c. Dashed line is 
the contribution from
$\Delta^{0}$-excitation, dotted from the background and
the solid  shows the sum.}
\end{figure}

\end{document}